\documentclass[
reprint,           
twocolumn,
superscriptaddress,
amsmath,           
amssymb,           
aps,               
prd,               
notitlepage,       
longbibliography,  
floatfix,          
nofootinbib,
]{revtex4-1}

\usepackage{amsmath}
\allowdisplaybreaks[4]
\usepackage{cancel}
\usepackage{extarrows}
\usepackage{tensor}     
\usepackage{float}
\usepackage[caption = false]{subfig} 
\usepackage[final]{graphicx}   
\usepackage[
colorlinks=true,        
citecolor=blue,         
linkcolor=blue,         
urlcolor=blue           
]{hyperref}             
\usepackage{bm}         
\usepackage{xcolor}     
\usepackage{lipsum}
\usepackage{color}      
\usepackage[utf8]{inputenc} 
\usepackage[section]{placeins} 
\usepackage{appendix}
\usepackage{units}
\newcommand{\nc}{\newcommand*}

\nc{\figurewidth}{3.2in}
\nc{\xbar}{\bar{x}}
\nc{\rhoeq}{\rho_{\mathrm{eq}}}
\nc{\zeq}{z_{\mathrm{eq}}}
\nc{\tla}{\tilde{\lambda}}
\nc{\dt}{\delta}
\nc{\Dt}{\Delta}
\nc{\vj}{\hat{j}}
\nc{\vl}{\hat{l}}
\nc{\hx}{\hat{x}}
\nc{\hy}{\hat{y}}
\nc{\bj}{\bm{j}}
\nc{\mJ}{\mathcal{J}}
\nc{\mP}{\mathcal{P}}
\nc{\Msun}{M_\odot}
\nc{\app}{\approx}
\nc{\av}[1]{\langle #1 \rangle}
\nc{\eq}[1]{Eq.~\eqref{#1}}
\nc{\al}{\alpha}
\nc{\Xstar}{X_{\ast}}
\nc{\seq}{\sigma_{\mathrm{eq}}}
\nc{\fpbh}{f_{\mathrm{pbh}}}
\nc{\vth}{\hat{\theta}}
\nc{\vla}{\hat{\lambda}}
\nc{\vd}{\hat{d}}
\nc{\Mmin}{M_{\mathrm{min}}}
\nc{\rmd}{\mathrm{d}}
\nc{\mmin}{{m_{\mathrm{min}}}}
\nc{\mmax}{{m_{\mathrm{max}}}}
\nc{\mR}{\mathcal{R}}
\nc{\tmR}{\tilde{\mathcal{R}}}
\nc{\s}{\sigma}
\nc{\ogw}{\Omega_{\mathrm{GW}}}
\nc{\addref}{[\textcolor{red}{add ref}] }
\nc{\Om}{\Omega}
\nc{\gm}{\gamma}
\nc{\Gm}{\Gamma}
\nc{\gpcyr}{\mathrm{Gpc}^{-3}\,\mathrm{yr}^{-1}}
\nc{\Eq}[1]{Eq.~\eqref{#1}}
\nc{\Fig}[1]{Fig.~\ref{#1}}
\nc{\Table}[1]{Table~\ref{#1}}
\nc{\lvc}{LIGO/Virgo} 
\nc{\Sec}[1]{Sec.~\ref{#1}}
\nc{\eg}{\textit{e.g.~}}
\nc{\SNR}{\mathrm{SNR}}
\nc{\bt}{\mathbf{t}}
\nc{\be}{\mathbf{\epsilon}}
\nc{\bn}{\mathbf{n}}
\nc{\bx}{\mathbf{x}}
\nc{\bk}{\mathbf{k}}
\nc{\bd}{\mathbf{d}}
\nc{\ba}{\mathbf{a}}
\nc{\bp}{\mathbf{p}}
\nc{\bnu}{\mathbf{\nu}}
\nc{\uni}{\mathrm{U}}
\nc{\logu}{\operatorname{\mathrm{log-U}}}
\nc{\RN}{\mathrm{RN}}
\nc{\BN}{\mathrm{BN}}
\nc{\GN}{\mathrm{GN}}
\nc{\mcN}{\mathcal{N}}
\nc{\GWB}{\mathrm{GW}}
\nc{\yr}{\mathrm{yr}}
\nc{\Am}{\mathcal{A}}
\nc{\Dm}{\mathcal{D}}
\nc{\Hm}{\mathcal{H}}
\nc{\sovast}{Soviet Ast.}
\nc{\hosc}{h_{\mathrm{osc}}}
\nc{\Posc}{\Psi_{\mathrm{osc}}}
\nc{\eV}{\mathrm{eV}}

\nc{\mrm}{\mathrm}
\nc{\BE}{B\scriptsize{AYES}\normalsize{E}\scriptsize{PHEM}\normalsize  }

\def\({\left(}
\def\){\right)}
\def\[{\left[}
\def\]{\right]}

\def\e{\begin{equation}}
\def\q{\end{equation}}
\def\m{\begin{eqnarray}}
\def\n{\end{eqnarray}}
\nc{\red}[1]{\textcolor{red}{#1}}
\begin{document}

\title{Constraining ultralight scalar dark matter couplings with the European Pulsar Timing Array second data release}     

\author{Yu-Mei Wu}
\email{wuyumei@yzu.edu.cn} 
\affiliation{Center for Gravitation and Cosmology,
College of Physical Science and Technology,
Yangzhou University, Yangzhou, 225009, China}
\affiliation{School of Fundamental Physics and Mathematical Sciences, Hangzhou Institute for Advanced Study, UCAS, Hangzhou 310024, China}
\affiliation{School of Physical Sciences, University of Chinese Academy of Sciences, No. 19A Yuquan Road, Beijing 100049, China}



\author{Qing-Guo Huang}
\email{Corresponding author: huangqg@itp.ac.cn}
\affiliation{School of Fundamental Physics and Mathematical Sciences, Hangzhou Institute for Advanced Study, UCAS, Hangzhou 310024, China}
\affiliation{CAS Key Laboratory of Theoretical Physics, 
    Institute of Theoretical Physics, Chinese Academy of Sciences,
    Beijing 100190, China}
\affiliation{School of Physical Sciences, 
    University of Chinese Academy of Sciences, 
    No. 19A Yuquan Road, Beijing 100049, China}


\begin{abstract}
Pulsar Timing Arrays (PTAs) offer an independent method for searching for ultralight dark matter (ULDM), whose wavelike nature induces periodic oscillations in the arrival times of radio pulses. In addition to this gravitational effect, the direct coupling between ULDM and ordinary matter results in pulsar spin fluctuations and reference clock shifts, leading to observable effects in PTAs. The second data release from the European PTA (EPTA) indicates that ULDM cannot account for all dark matter in the mass range $m_{\phi} \in [10^{-24.0}, 10^{-23.3}] \text{ eV}$ based solely on gravitational effects. In this work, we derive constraints on the coupling coefficients by considering both gravitational and coupling effects. Our results demonstrate that EPTA provides stronger constraints on these couplings than previous PTA experiments, and it establishes similar or even tighter constraints compared to other precise experiments, such as atomic clock experiments.
\end{abstract}


	
\maketitle
	
	
\section{Introduction}

The existence of dark matter has been confirmed by numerous experiments, including galaxy rotation curves \cite{1980ApJ...238..471R,1982ApJ...261..439R}, velocity dispersions \cite{1976ApJ...204..668F}, and gravitational lensing \cite{Massey:2010hh}. Additionally, observations of the cosmic microwave background suggest that dark matter constitutes roughly five times more of the universe's mass than visible matter \citep{Planck:2018vyg}, according to the widely accepted Lambda cold dark matter paradigm. However, The fundamental nature of dark matter, including its composition and interaction with the Standard Model, is still a mystery.
Among the many candidates for cold dark matter, weakly interacting massive particles (WIMPs) were initially highly anticipated, yet extensive searches have so far yielded no positive results \cite{Schumann:2019eaa}. Moreover, traditional CDM candidates encounter challenges on small scales, such as the discrepancy between predictions of steep central densities in galaxy cores and observed flatter density profiles (the cusp-core problem, \cite{Gentile:2004tb,deBlok:2009sp}), as well as the mismatch between the predicted steep increase in the number of low-mass subhalos and the significantly lower number of observed satellites around the galaxies (the missing satellite problem, \cite{Moore:1999nt,Klypin:1999uc}). One potential solution to these issues is the hypothesis that dark matter consists of ultralight bosons \citep{Hu:2000ke}, typically with mass $m_{\phi} \ll \unit[10^{-6}]{\eV}$, whose large de Broglie wavelength can efficiently suppress structure formation on small scales \citep{Hui:2016ltb}.

The phenomenon of ultralight dark matter (ULDM) encompasses a broad range of possibilities, allowing it to be constrained by various observational data. For example, measurements of the cosmic microwave background (CMB) exclude ULDM with masses below $m_{\phi}<\unit[10^{-24}]{\eV}$ \cite{Hlozek:2017zzf}. Dwarf galaxy dynamics \cite{Marsh:2018zyw} provide a stronger lower bound at $m_{\phi}>\unit[10^{-22}]{\eV}$, while galaxy rotation curves \cite{Bar:2018acw} and Lyman-$\alpha$ forest observations \cite{Rogers:2020ltq} further tighten this limit to $m_{\phi}>\unit[10^{-21}]{\eV}$. However, these constraints from non-CMB experiments are heavily reliant on models of small-scale structure formation \cite{Schive:2014hza,Zhang:2017chj}, which introduce considerable uncertainties. To overcome these uncertainties, it is crucial to employ complementary methods that can provide independent constraints on ULDM in this mass range. One promising approach involves pulsar timing arrays (PTAs), which monitor the highly regular radio pulses of millisecond pulsars over extended periods \cite{1978SvA....22...36S,Detweiler:1979wn,1990ApJ...361..300F}. As sensitive detectors operating in the nanohertz frequency range, PTAs offer a feasible means to detect or constrain ULDM with masses between $\unit[10^{-24}]{\eV}$ and $\unit[10^{-20}]{\eV}$. 

ULDM can generate detectable signals in PTAs through two primary mechanisms. First, the oscillations of ULDM can directly induce spacetime fluctuations, resulting in changes to the times of arrival (ToAs) of radio pulses \cite{Khmelnitsky:2013lxt, Nomura:2019cvc, Wu:2023dnp}, known as timing residuals—the differences between the actual measured and predicted ToAs. This phenomenon is referred to as the gravitational effect of ULDM. Second, ULDM may interact with the Standard Model, and its coupling with the pulsars and the Earth, both composed of ordinary matter, can leave observable imprints on the ToAs in a more subtle manner \cite{PPTA:2021uzb,Armaleo:2020yml,Kaplan:2022lmz}. This interaction is referred to as the coupling effect. 
Several studies have explored ULDM using PTA datasets, but the gravitational and coupling effects are typically considered in isolation \cite{Porayko:2018sfa,PPTA:2021uzb,PPTA:2022eul,EuropeanPulsarTimingArray:2023egv,NANOGrav:2023hvm}. Furthermore, analyses of the coupling effect often assume that ultralight fields constitute the entirety of dark matter. However, two critical points warrant emphasis. First, the gravitational effect of ULDM is always present, regardless of any potential interaction with the Standard Model. Second, recent constraints on ULDM density, derived solely from the gravitational effect using data from the European PTA second data release (EPTA DR2), have ruled out the possibility that ultralight particles constitute $100\%$ of the local dark matter within the mass range of $[10^{-24}, 10^{-23.3}]$ eV \cite{EuropeanPulsarTimingArray:2023egv}. This underscores the importance of considering both effects in the analyses of coupling constraints. 

In this paper, we adopt a more comprehensive approach by simultaneously considering both the gravitational and coupling effects based on EPTA DR2, and we present updated constraints for the ULDM couplings derived from this unified analysis. The structure of this work is as follows: \Sec{sec2} introduces the observable pulsar timing effects induced by scalar dark matter, \Sec{sec3} provides details of our data analysis, while \Sec{sec4} present the results and conclusions.

\section{Gravitational effect and coupling effect from scalar dark matter}\label{sec2}

In this section, we provide a brief overview of the timing signal induced by scalar field dark matter, focusing on both the gravitational and coupling effects. For a more detailed derivation, one can refer to \cite{Khmelnitsky:2013lxt,Kaplan:2022lmz}.

On the galactic scale, the spacetime can be approximately described by Minkowski. The Lagrangian  of a free scalar dark matter field with mass $m_{\phi}$ is 
\e
L_{\phi}=\frac{1}{2}\partial^{\mu}\phi\partial_{\mu}\phi-\frac{1}{2}m_{\phi}^2\phi^2.
\q
As the characteristic speed of the dark matter is non-relativistic $v\sim 10^{-3}$, and the occupation number of the scalar field is huge, the ultralight scalar dark matter can be described as a monochromatic classical wave,
\e
\phi(\bx,t)=A \hat{\phi}(\bx)\cos(2\pi f t+\alpha(\bx)),
\q
where the overall oscillation amplitude $A$ is related to the energy density of the scalar field $\rho_{\phi}$, the oscillation frequency $f$ is determined by the particle mass $m_{\phi}$,
\e
A=\frac{\sqrt{2\rho_{\phi}}}{m_{\phi}}, \qquad
f=\frac{m_{\phi}}{2\pi},
\q
$\alpha(\bx)$ is the position-dependent phase, and the additional random factor $\hat{\phi}(\bx)$ is introduced to capture the stochastic behavior when accounting for the interference among the field modes with different frequencies due to the small velocity dispersion \cite{Centers:2019dyn}. 

The oscillating scalar field would contribute to a time-dependent tensor-momentum tensor, which induces a gravitational potential perturbation to the flat background. The radio pulse travel from the pulsar to the Earth would experience a redshift and hence a delay on the arrival time,  
\e
\Delta t_{grav}=\frac{\Phi_c}{2m_{\phi}} \[\hat{\phi}_E^2\sin(2m_{\phi}t+2\gm_{E})-\hat{\phi}_P^2\sin(2m_{\phi}t+2\gm_{P})\],
\label{GE}
\q
which is referred to the gravitational effect of ULDM.
Here, $\Phi_c$ is the amplitude of oscillating gravitational potential,
\e
\Phi_c=\frac{\pi G \rho_{\phi}}{m_{\phi}^2}\approx 6.1\times 10^{-18} \(\frac{m_{\phi}}{10^{-22}eV}\)^{-2}\(\frac{\rho_{\phi}}{\rho_0}\),
\q
with $\rho_0=\unit[0.4]{\rm{GeV/cm^3}}$ denoting the local DM density, $\hat{\phi}_E=\hat{\phi}(\bx_{E})$ and $\hat{\phi}_P=\hat{\phi}(\bx_{P})$ are the oscillation amplitude normalized factors of the scalar field at the Earth and the pulsar, respectively, while $\gm_{E}=\alpha(\bx_{E})$ and $\gm_{P}=\alpha\left(\bx_{P}\right)-m_{\phi}\left|\mathbf{x}_{P}\right|$ represent the redefined phases at these locations.

Apart from the gravitational effect inherent in any matter in motion, the ULDM could couple to Standard Model in a plethora of ways. The possible coupling to the QED sector and QCD sector can be parameterized as \cite{Damour:2010rp},
\m
L_{int}&&=\frac{\phi}{\Lambda}\left(\frac{d_{\gm}}{4e^2} F_{\mu\nu}F^{\mu\nu}-\sum_{f=e,\mu} d_f m_f \bar{f}f \right. \notag \\
&&\left.+ \frac{d_g\beta_3}{2g_3}G_{\mu\nu}^A G_{A}^{\mu\nu}-\sum_{q=u,d}(d_q+\gm_q d_g)m_g \hat{q}q\right),
\label{Lint}
\n
where $\Lambda = M_{pl}/\sqrt{4\pi}$, with $M_{pl}$ denoting the Planck mass. Here, $F_{\mu\nu}$ represents the electromagnetic field strength tensor, while $G_{\mu\nu}^{A}$ is the gauge-invariant gluon strength tensor. The QCD gauge coupling is indicated by $g_3$, and $\beta_3$ refers to the beta function that governs the running of $g_3$. Additionally, $\gamma_{q}$ represents the light quark anomalous dimension, and $m_f$ and $m_q$ denote the masses of fermions and quarks, respectively. The $d$ values are dimensional coupling coefficients. Through these couplings, fluctuations in the ULDM background induce variations in the fundamental constants of the Standard Model, including the electromagnetic coupling constant and the masses of electrons, muons, quarks, and nucleons,
\m
&& \frac{\dt \alpha}{\alpha}=\frac{d_{\gm}}{\Lambda}\phi, \qquad \quad \,\,\,\, \frac{\dt m_{e,\mu}}{m_{e,\mu}}=\frac{d_{e, \mu}}{\Lambda}\phi, \notag \\
&&\frac{\dt m_{q}}{m_{q}}=\frac{d_{q}}{\Lambda}\phi,
\qquad \quad \frac{\dt m_{p,n}}{m_{p,n}}=\frac{d_{g}+C_n d_{\hat{m}}}{\Lambda}\phi,
\n
where  $C_n=0.048$, and $d_{\hat{m}}$ is a symmetric combination of the quark mass coupling  $d_{\hat{m}}=(d_u m_u +d_d m_d)/(m_u+m_d)$.

The timing residuals resulting from fluctuations in fundamental constants arise from two main aspects. First, shifts in particle mass lead to variations in the pulsar's moment of inertia, which in turn cause fluctuations in the pulsar's spin frequency by the conservation of angular momentum.  Second, the atomic clock on Earth, used to measure the ToAs of radio pulses, also experiences frequency shifts, contributing to the observed timing residuals. Overall, the timing residuals induced by the coupling effect, stemming from both pulsar spin fluctuations and shifts in the Earth’s atomic clock, are expressed as:
\begin{small}
\m
\Delta t_{coupl}
=\frac{A_i}{m_{\phi}}\[y_{P}^i \hat{\phi}_P\sin(m_{\phi}t+\gm_{P})+y_{E}^i \hat{\phi}_E\sin(m_{\phi}t+\gm_{E})\], \notag\\
\label{CE}
\n
\end{small}
where $A_i=d_i \sqrt{2\rho_{\phi}}/(m_{\phi} \Lambda)$, $y_{P}$ and $y_{E}$ respectively parameterize the pulsar or the the atomic clock sensitivity to a specific coupling and are specifically given by
\m
&& y_{P}^g=-5,\quad y_{P}^{\hat{m}}=-0.24,\quad y_{P}^{\gm}=0, \notag \\
&&y_{P}^{\mu}=2\times 10^{-3},\quad y_{P}^{e}=1.7\times 10^{-5} \, ;\notag \\
&& y_{E}^{g}=1,\quad y_{E}^{\hat{m}}= 0.296,\quad y_{E}^{\gm}=4.83, \notag\\
&&y_{E}^{\mu}=0,\quad y_{E}^{e}=2.
\n

\section{Data analysis}\label{sec3}

We employ the EPTA DR2 dataset to constrain the coupling effect of ULDM. This dataset includes observational data from 25 pulsars, covering a total observation period of 24.7 years \cite{EPTA:2023sfo}. The official EPTA collaboration has previously utilized this dataset to investigate the gravitational effect of ULDM \cite{EuropeanPulsarTimingArray:2023egv}, and we will adopt a similar methodology in our analyses.

The expected arrival times of radio pulses for each individual pulsar are described by a timing model that incorporates various pulsar characteristics, such as its position in the sky, spin-down rate, and proper motion. Timing residuals are obtained by subtracting these expected arrival times from the actual observed times. These residuals arise from several contributing factors, including inaccuracies in the timing model, stochastic noise during the generation and propagation of the radio pulses, and the potential astrophysical signals of interest that we aim to detect. The timing models for each pulsar are included with the dataset. To effectively extract the desired signals, a comprehensive analysis of the noise present in the data is essential.

The stochastic noise specific to each pulsar can be classified into two categories: time-uncorrelated white noise and time-correlated red noise. White noise accounts for measurement uncertainties and other potential time-independent errors, which may arise from factors such as instrumental miscalibration and changes in pulse profiles \cite{Lentati:2015qwp, NANOGrav:2015qfw}. This white noise is modeled using two parameters: “EFAC,” which scales the ToA uncertainties, and “EQUAD,” an additional term incorporated in quadrature. In contrast, red noise includes achromatic spin noise intrinsic to each pulsar, resulting from rotational instabilities\cite{Shannon:2010bv}, as well as chromatic dispersion variations occurring as radio pulses traverse the interstellar medium \cite{Keith:2012ht}. Red noise is typically modeled by a power-law spectrum characterized by two parameters: the amplitude $A_{\text{red}}$ and the spectral index $\gamma_{\text{red}}$.
Furthermore, in the search for stochastic gravitational-wave background (SGWB), while the critical detection criterion—the distinctive Hellings-Downs quadrupole correlation—has not yet yielded strong evidence, a common uncorrelated spectrum across all pulsars, likely arising from the auto-correlation component of the SGWB, has shown considerable promise \cite{EPTA:2023fyk}. Therefore, we will also incorporate the contribution of this common uncorrelated red noise when modeling the timing residuals.

The signals from ULDM, arising from both the gravitational and coupling effects, are described by \Eq{GE} and \Eq{CE} as deterministic sinusoidal functions, characterized by amplitude parameters $\Phi_c$ and $A_i$, amplitude normalized factors $\hat{\phi}_E$ and $\hat{\phi}_P$, mass (or frequency) parameter $m_{\phi}$, and phase parameters $\gamma_{E}$ and $\gamma_{P}$. For scales smaller than the de Broglie wavelength $l \sim \unit[0.4]{\rm{kpc \,(10^{-22}eV/m_{\phi})}}$, the ULDM scalar field demonstrates coherent oscillations with a consistent amplitude factor $\hat{\phi}$. Given that the typical distances between pulsars and between the pulsars and Earth are comparable to this coherence length, we consider two representative scenarios: ``Correlated," where the Earth and all pulsars share the same normalized amplitude factor ($\hat{\phi}_E = \hat{\phi}_P$), and ``Uncorrelated," where the Earth and pulsars have independent factors. To facilitate a more flexible search for or constraint on ULDM signals, we partition the mass range $m_{\phi} \in [10^{-24}, 10^{-22}] \eV$ into equally spaced small segments. Additionally, since ULDM cannot exceed the total dark matter density (i.e., $\rho_{\phi} < \rho_{0}$), we impose an upper limit $\Phi_0$ on the amplitude $\Phi_c$.

The analysis of PTA data begins with a thorough examination of the noise for each individual pulsar. We then fix the white noise at their maximum likelihood values to reduce the computational cost. The red noises, which can be specific to individual pulsars or common across all pulsars, are modeled simultaneously with the target signal. All relevant parameters and their associated priors are outlined in Table 1. We will utilize Bayesian methodology to assess the presence or absence of the signal by calculating the Bayes factor $\mathbf{BF}$ between the ``noise + signal" model and the ``noise only" model. A positive indication of the signal's presence can be declared when $\ln \mathbf{BF} >3$. If no signal is detected, we will estimate the $95\%$ upper limit for the amplitude. To evaluate the likelihood, we utilize the \texttt{enterprise} \citep{enterprise} and \texttt{enterprise\_extension} \citep{enterprise_extensioins} packages, applying the product-space method \citep{10.2307/2346151, Hee:2015eba, Taylor:2020zpk} to compute the Bayes factor. For parameter estimation, we implement Markov-chain Monte Carlo sampling using the \texttt{PTMCMCSampler} package \citep{justin_ellis_2017_1037579}.

\begin{table*}[!htbp]
	\footnotesize
	\caption{Parameters and their prior distributions used in the analyses. "U" and "log-U" denote uniform and log-uniform distributions, respectively. "One parameter for PTA" indicates that the parameter is shared across the entire data set, while "one parameter per pulsar" signifies that the parameter varies from pulsar to pulsar.}
	\label{prior}
	\begin{tabular}{c c c c}
		\hline
		\textbf{Parameter} & \textbf{description} & \textbf{prior} & \textbf{comments} \\
		\hline
		\multicolumn{4}{c}{White noise}\,\\	        
		$E_{k}$ & EFAC per backend/receiver system & $\uni[0, 10]$ & single-pulsar analysis only \\
		$Q_{k}\,[\mrm{s}]$ & EQUAD per backend/receiver system & $\logu[-8.5, -5]$ & single-pulsar analysis only \\
		\hline
		\multicolumn{4}{c}{Red noise (including SN and DM)} \\
		$\Am_{\RN}$ & Red-noise power-law amplitude &$\logu[-20, -8]$ & one parameter per pulsar\, \\
		$\gamma_{\RN}$ &red-noise power-law index  &$\uni[0,10]$ & one parameter per pulsar\, \\
		\hline
		\multicolumn{4}{c}{Common noise} \\
		$\Am_{\rm{GWB}}$ & common-noise power-law amplitude &$\logu[-20, -6]$ & one parameter for PTA\, \\
		\hline
		\multicolumn{4}{c}{Ultralight dark matter signal}\,\\
		$\Phi_{c}$ & ULDM gravitational effect amplitude &$\logu[-20, \log_{10}\Phi_{0}]$ & one parameter for PTA\, \\
		$A_{i}$ & ULDM coupling effect amplitude &$\logu[-20, -10]$ & one parameter for PTA\, \\
		$m_{\phi}$ & ULDM mass &$\logu[-24, -22]$ & one parameter for PTA\, \\
		$\hat{\phi}_{E}^2$ & Earth normalized signal amplitude &$e^{-x}$ & one parameter for PTA\, \\
		$\hat{\phi}_{P}^2$ & Pulsar normalized signal amplitude &$e^{-x}$ & one parameter per pulsar\, \\
		$\gm_E$ & oscillation phase on Earth &$\uni[0,2\pi]$ & one parameter for PTA\, \\
		$\gm_P$ & equivalent oscillation phase on pulsar &$\uni[0,2\pi]$ & one parameter per pulsar\, \\
		\hline
	\end{tabular}
\end{table*}


\section{results and discussion}\label{sec4}

\begin{figure*}[htbp!]
	\centering
	\includegraphics[width=1.\textwidth]{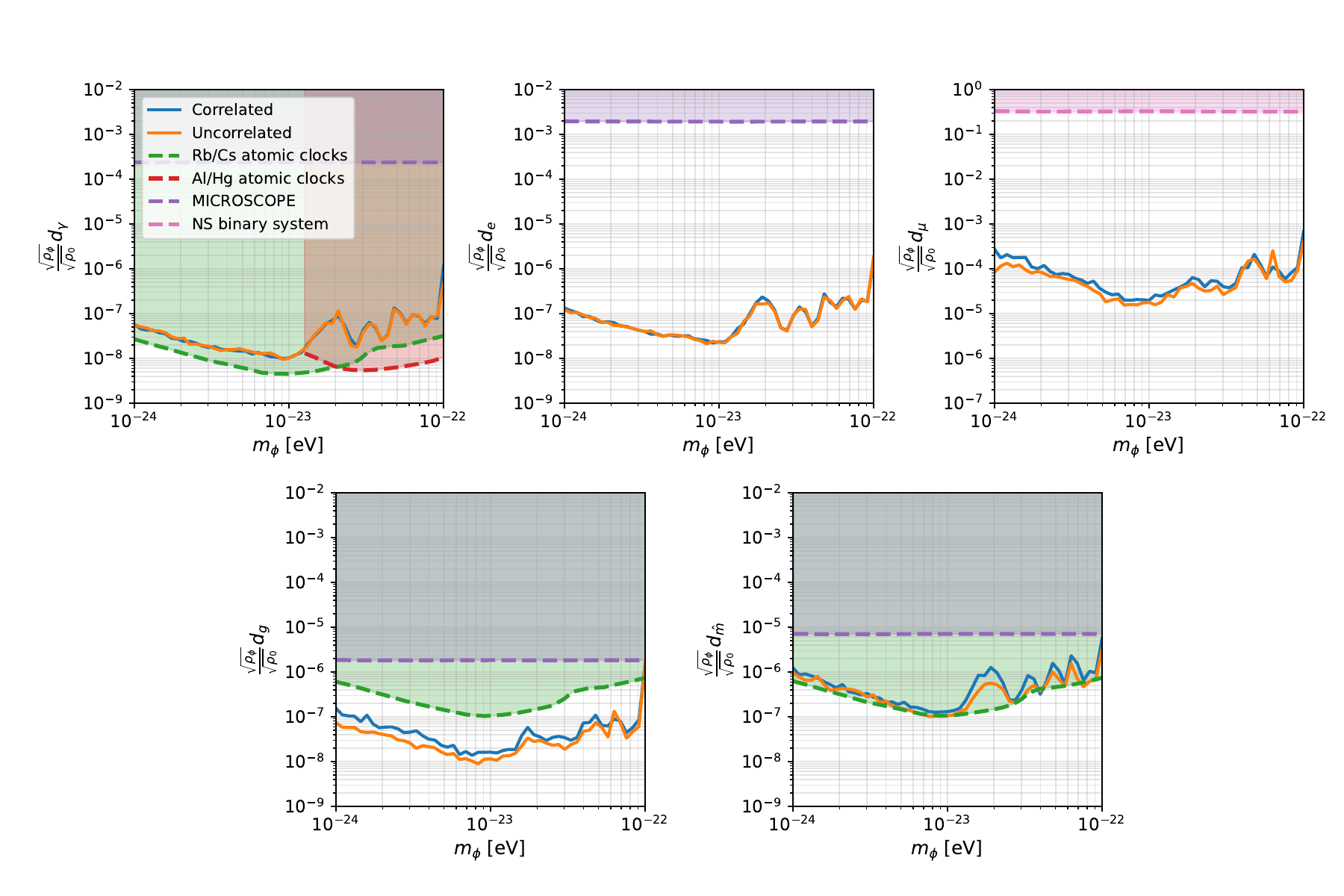}\caption{Upper limits on the five dimensionless parameters $d_{i}\sqrt{\rho_{\phi}}/\sqrt{\rho_{0}}$ (solid lines) as a function of mass, derived from the coupling effect of the ULDM. The blue and orange solid lines represent searches for the ULDM signal in correlated and uncorrelated scenarios, respectively. Dashed lines indicate constraints from other experiments, with green, red, purple, and pink dashed lines corresponding to the “Rb/Cs atomic clocks”\cite{Hees:2016gop}, “Al/Hg atomic clocks” \cite{BACON:2020ubh}, “MICROSCOPE” \cite{Berge:2017ovy} and “NS binary system” \cite{KumarPoddar:2019ceq,Dror:2019uea} experiments, respectively.\label{coupling}}
\end{figure*}

In our analysis of the coupling effect described by \Eq{CE}, we consider the induced timing signals from individual couplings with the photon ($d_{\gamma}$), electron ($d_e$), muon ($d_{\mu}$), gluon ($d_{g}$), and quark ($d_q$) separately, alongside the ubiquitous gravitational effect outlined in \Eq{GE}. Within the mass range we explore, $10^{-24} \eV < m_{\phi} < 10^{-22} \eV$, we found no positive evidence for the ULDM signal, as indicated by $\ln \mathbf{BF} < 1$. Consequently, we compute constraints on the ULDM coupling constants. It is important to note that the amplitude parameters $A_i$ encompass both the coupling constants $d_i$ and the ULDM density $\rho_{\phi}$; thus, we can only derive upper limits for the product $d_i \sqrt{\rho_{\phi}}$ based on the 95th percentile of the marginalized posterior distributions of $A_i$. The results of these constraints are presented in \Fig{coupling}.

The constraints on individual coupling constants from the ULDM coupling effect are displayed in each panel for both ``Correlated" and ``Uncorrelated" scenarios, represented by solid lines. It is worth noting that the parameters $d_{\gamma}$ and $d_e$ couple significantly more strongly to the Earth term than to the pulsar term, resulting in minimal differences between the ``Correlated" and ``Uncorrelated" cases. For comparison, we also include other experimental constraints, indicated by dashed lines. Our findings reveal that PTA experiments are slightly less sensitive than the most precise atomic clock results \cite{Hees:2016gop,Kennedy:2020bac} in constraining the coupling constant $d_{\gamma}$, yet they are roughly comparable for $d_{\hat{m}}$ and exhibit greater sensitivity for $d_g$. Importantly, PTAs can constrain the $d_e$ coupling, which is not feasible with atomic clock experiments due to the cancellation of the $d_e$ dependence in the relative frequency shifts between clocks that use the same atomic transition.  Furthermore, PTAs surpass Weak Equivalent Principle constraints \cite{Schlamminger:2007ht,Berge:2017ovy} by several orders of magnitude. While there are currently no laboratory constraints on $d_{\mu}$ due to the scarcity of muons on Earth, neutron stars, which host a substantial number of muons, provide a more favorable environment for constraining $d_{\mu}$. Consequently, PTAs are exceptionally sensitive to $d_{\mu}$, enhancing the previous projected constraints from the orbital decay of a neutron star (NS) binary system \cite{KumarPoddar:2019ceq,Dror:2019uea} by three orders of magnitude.

In addition, our constraints on $d_i\sqrt{\rho_{\phi}}$ based on EPTA DR2 are $2 \sim 3$ times stricter than those from the NANOGrav collaboration's 15-year datasets \cite{NANOGrav:2023hvm}, which is consistent with the comparison of constraints on dark matter density $\rho_{\phi}$ derived from pure gravitational effects in previous searches by both collaborations \cite{EuropeanPulsarTimingArray:2023egv,NANOGrav:2023hvm}. This suggests that EPTA exhibits higher sensitivity to deterministic signals due to its longer observational span. 

In the future, as sensitivity improves, if the dataset constrains the ULDM to a fraction $f_{\phi}$ of the total dark matter abundance, the constraints on the coupling coefficients $d_i$ could tighten by a factor of $\sqrt{f_{\phi}}$. Additionally, it may become possible to detect a sinusoidal monochromatic signal in PTAs. However, such a signal could originate from various sources, including the ULDM gravitational effect, the ULDM coupling effect, or a continuous gravitational wave. In this case, we will need to utilize the correlation of signals between pulsars to resolve their degeneracy \cite{Kaplan:2022lmz,Omiya:2023bio,Cai:2024thd}.

\begin{acknowledgments}
This work is supported by the grants from NSFC (Grant No.~12250010, 11991052), Key Research Program of Frontier Sciences, CAS, Grant No.~ZDBS-LY-7009. We acknowledge the use of HPC Cluster of ITP-CAS.
\end{acknowledgments}
\bibliography{./ref}

\end{document}